\newcommand{\diracslash}[1]{#1\llap{/\kern2pt}}

\newcommand{\be}{\begin{equation}}
\newcommand{\ee}{\end{equation}}
\newcommand{\bea}{\begin{eqnarray}}
\newcommand{\eea}{\end{eqnarray}}
\newcommand{\ba}[1]{\begin{array}{#1}}
\newcommand{\ea}{\end{array}}

\documentclass[twocolumn,superscriptaddress,secnumarabic,amssymb,nobibnotes,nofootinbib,aps,prl,showpacs]{revtex4}
\usepackage{graphicx}
\begin{document}
\setlength{\topmargin}{-0.05in}

\title{Interior gap superfluidity
in  a two-component Fermi gas of  atoms}

\author{Bimalendu Deb}
 \affiliation{ Physical
Research Laboratory, Navrangpura, Ahmedabad 380 009, India}%

\author{Amruta Mishra}
\altaffiliation[Present address: ]{Physics Department, Indian Institute of Technology, Delhi, New Delhi 110 016, India.}%
\affiliation{Institute f\"ur Theoretische Physik, Universit\"at Frankfurt, D-60054 Frankfurt, Germany }%

\author{Hiranmaya Mishra}
\affiliation{ Physical Research Laboratory, Navrangpura, Ahmedabad
380 009, India}

\author{Prasanta K. Panigrahi}
 \affiliation{
Physical Research Laboratory, Navrangpura, Ahmedabad 380 009,
India}
 \affiliation{
School of Physics, University of Hyderabad, Hyderabad-500 046,
India}


\def\be{\begin{equation}}
\def\ee{\end{equation}}
\def\bearr{\begin{eqnarray}}
\def\eearr{\end{eqnarray}}
\def\zbf#1{{\bf {#1}}}
\def\bfm#1{\mbox{\boldmath $#1$}}
\def\hf{\frac{1}{2}}

\begin{abstract}
A new superfluid phase in Fermi matter, termed as ``interior gap"
(IG) or ``breached pair",  has been  recently  predicted by Liu
and Wilczek [Phys.Rev.Lett. {\bf 90}, 047002 (2003)]. This results
from pairing between fermions of two species  having essentially
different Fermi surfaces. Using a nonperturbative variational
approach,  we analyze the features,  such as energy gap, momentum
distributions, and elementary  excitations associated with the
predicted phase. We discuss possible realization of this phase in
two-component Fermi gases in an optical trap.
\end{abstract}

\pacs{03.75.kk,74.20.-z,32.80.Lg}

\maketitle

 Since the first realization
of Fermi degeneracy in  $^{40}$K gas by Jin's group  \cite{jin},
several other groups \cite{huletf,solomon,thomas,
ketterlef,inguscio} have also  achieved
 quantum degeneracy in  atomic Fermi gases.  A
recent experiment by  Regal {\it et al.}  \cite{regal}, has
reported condensation of fermionic atom-pairs in trapped $^{40}$K
gas. Recently, several  groups  have  demonstrated formation of
cold molecules \cite{coldmol,salomonnew} and molecular
Bose-Einstein condensates \cite{molecules} from trapped Fermi
gases of atoms near the Feshbach resonance.  O'Hara  {\it et al.}
\cite{sf} demonstrated strongly interacting degenerate Fermi gas
of $^6$Li atoms. Studies on trapped degenerate Fermi gases are
important in the context of diverse fields, such as nuclear
physics, astrophysics, strong-coupling superconductivity and
superfluidity, \cite{resonant} and so on.

A new form of Fermi matter, associated with what is termed as
``interior gap " (IG) superfluidity \cite{wilczek}, has recently
come into fore. Liu and Wilczek \cite{wilczek} have shown that, in
two  species of fermions differing in Fermi momenta,  an
attractive interaction can lead to pairing  within the interior of
the larger Fermi surface, while the exterior remains gapless. This
pairing phenomenon gives rise to IG superfluidity having
superfluid and normal Fermi liquid components simultaneously. Many
years ago, an analogous state of two-species fermions was
predicted by Sarma \cite{sarma}, which, however, corresponded to a
metastable phase. Recently, it was shown in the context of color
superconductivity that, such a phase is the only stable phase when
the relative density of the two species \cite{wil2,yip} is kept
fixed. IG superfluidity  arises naturally in finite density quark
matter \cite{igor,mishra,wil2,rajgopal}.

 In this
paper, we analyze the energy gap, momentum distributions and
quasiparticle excitations in  an  IG superfluid system. For
possible realizations of this phase, optically trapped
two-component atomic Fermi gases may be useful. It has already
been  suggested \cite{wilczek} that a mixture of $^6$Li and
$^{40}$K Fermi gases should be used to create  this new phase. Let
us consider a mixture of two hyperfine spin components,  for
instance, $|1\rangle = |F=1/2,m_F=1/2\rangle$ and $|2\rangle =
|F=1/2,m_F=-1/2\rangle$ of $^6$Li atoms,  in an optical trap.  The
relative number density of the two components can be controlled by
a rf field \cite{sf,rf,zwierlein}. The lifetime of each  spin
component is long enough \cite{stable} to carry out
 an experiment at a fixed  relative density which
is a necessary condition for IG superfluidity \cite{wilczek}. It
may be noted that controlled mixtures, other than 50:50, of
fermion gas have already been created \cite{zwierlein}.

To discuss IG superfluidity, let us consider  a system of
interacting fermions of two species in a harmonic trap, described
by the Hamiltonian $\hat{H} = \hat{H}_0  + \hat{H}_I$ where
\begin{equation}
\hat{H}_0 = \sum_{i=1,2} \int d{\mathbf r}
\hat{\Psi}_i^{\dagger}({\mathbf r}) \left [
-\frac{\hbar^2}{2m_i}\nabla^2 + V_{ho}({\mathbf r}) -
\mu_{i}({\mathbf r}) \right ]\hat{\Psi}_i({\mathbf r}),
\end{equation}
where $V_{ho}({\mathbf r})$ is the harmonic trapping potential and
 \begin{equation}
 \hat{H}_I = \frac{1}{2} \int \int d{\mathbf r} d{\mathbf r}^{\prime}
\hat{\Psi}_1^{\dagger}({\mathbf r})\hat{\Psi}_2^{\dagger}({\mathbf
r}')V(|{\mathbf r}-{\mathbf r}'|)\hat{\Psi}_1({\mathbf
r})\hat{\Psi}_2({\mathbf r}').
\end{equation}
Here, $\hat{\Psi}_i({\mathbf r})$ represents fermionic field
operator, $\mu_i({\mathbf r})$ denotes  the  local chemical
potential and $V(|{\mathbf r}-{\mathbf r}'|)$ is the two-body
interaction potential.  We now introduce the annihilation operator
$\hat{c}_i({\mathbf k})$ which, for a uniform gas, is related to
field operator by $\hat{\Psi}_i({\mathbf r}) = (1/\sqrt{V})
\sum_{{\mathbf k}} \exp(-i{\mathbf k}.{\mathbf
r})\hat{c}_i({\mathbf k})$, where $V$ is the volume of the system.
For a uniform gas with number density $n$, the mean field energy
in the weak interaction regime is $U_{mf} = gn$, where $g =
4\pi\hbar^2a_s/(2\tilde{m})$, with  $\tilde{m} = m_1m_2/(m_1+m_2)$
being the reduced mass, and $a_s$ the s-wave scattering length. In
the strong interaction regime, it is unitarity-limited and takes
the form $U_{mf} = \beta E_{F}$ \cite{sf,heiselberg,unitary},
where $\beta$ is a constant. In this regime, one can define an
effective scattering length from the relations $ U_{mf} = gn =
\beta E_F$ yielding $a_{{\rm eff}} \propto 1/k_F$. Under local
density approximation (LDA) in a harmonic trap of potential
$V_{ho}$, the equilibrium conditions are given by \cite{houbiers}
\begin{eqnarray}
\frac{\hbar^2}{2m_i}[6\pi^2n_i({\mathbf r})]^{2/3} +
g|\epsilon_{ij}|n_j({\mathbf r}) + V_{ho}({\mathbf
r})=\mu_i({\mathbf r}),
\end{eqnarray}
where  $ \epsilon_{ij} $ is the Levi-Civita tensor. LDA is
particularly good when the local Fermi energy is larger than the
average level spacing of the trap and the coherence length of the
fermion pair is shorter than the average trap size
\cite{houbiers}. The Fermi momentum of the $i$th component is then
related to the respective peak density $n_i({\mathbf r}_0)$ at
trap center by $k_{F_i} = [6\pi^2n_i({\mathbf r}_0)]^{1/3}$. In
what follows, we thus consider a uniform gas and derive the gap
equation.

To consider the pairing state, we take the variational ansatz as
given by \cite{mishra,npa}
\begin{equation}
|\Omega\rangle=\exp\left [ \frac{1}{2}\int \left(\hat{c}_{i}(\zbf
k)^\dagger
 f(\zbf k) \hat{c}_{j}(-\zbf k)^\dagger
\right) \epsilon_{ij}d\zbf k -{\mathrm H.c.}\right] |0\rangle,
\label{omg}
\end{equation}
where $|0\rangle$ represents the vacuum, annihilated by
$\hat{c}_i$'s. To include the effects of temperature and density,
we write down the state at finite temperature and density
$|\Omega(\beta,\mu)\rangle$ taking a thermal Bogoliubov
transformation over the state $|\Omega\rangle$ using thermofield
dynamics  as described in Refs. \cite{tfd,amph4}. We then have,
\begin{equation}
|\Omega(\beta,\mu)\rangle=\exp\left( \int c_i' (\zbf k)^\dagger
\theta_i(\zbf k, \beta) \tilde{c}_i' (-\zbf k)^\dagger
 d\zbf k - {\mathrm H.c.}\right)
|\Omega\rangle. \label{bth}
\end{equation}
The superscript ``prime" refers to operators in $|\Omega\rangle $
basis, i.e., $c_i' |\Omega \rangle =0$. The tilde operators are
the ones in the extended Hilbert space. In Eq. (\ref{bth}), the
function $\theta_i(\zbf k,\beta,\mu)$ will be related to the
distribution functions, for the two species. Using the Bogoliubov
transformation, it is easy to calculate the thermodynamic
potential given by \be \Omega=\langle\Omega(\beta,\mu)|{\cal
H}-\mu_j\psi_j^\dagger\psi_j
|\Omega(\beta,\mu)|\rangle-\frac{1}{\beta} s; \label{omg1} \ee
where ${\cal H}$ is the Hamiltonian density for constant
potential. In the above  \cite{tfd},  $s =
-\frac{2}{(2\pi)^3}\sum_{i}\int d \zbf k \Big ( \sin^2
\theta_{i}\ln \sin^2 \theta_{i} +\cos^2 \theta_{i}\ln \cos^2
\theta_{i}\Big ) $ is the entropy density.

Minimizing the thermodynamic potential $\Omega$ with respect to
condensate function $f (\zbf k)$, we get $\tan 2f(\zbf k)=- (2g
I_D)/(\epsilon_1-\mu_1+g\rho_2 +\epsilon_2-\mu_2+ g\rho_1)\equiv
\Delta/(\bar \epsilon-\bar \nu)$, where we have defined the gap
$\Delta=-g I_D$. The order parameter here is $\langle\psi_1(\zbf
x) \psi_2(\zbf x)\rangle=-I_D$ where,
$I_D=\frac{1}{2(2\pi)^3}\int{\sin 2f(\zbf k)}\left(1- \sin^2
\theta_1-\sin^2 \theta_2\right)d \zbf k.$ Further,
$\bar\epsilon=(\epsilon_1+\epsilon_2)/2$, $\bar\nu=
(\nu_{1}+\nu_{2})/2$, $\nu_{i}$ is the {\em interacting} chemical
potential given as $\nu_{i}=\mu_{i}-{g}\rho_j|\epsilon_{ij}|$.
Thus, it may be noted that the pair condensate functions depend
upon the {\em average} energy and the {\em average} chemical
potential of the fermions that condense. Finally, the minimization
of the thermodynamic potential with respect to the thermal
functions $\theta_{i}(\zbf k)$
 yields $\sin^2\theta_{i}={1}/({\exp(\beta\omega_{i})+1})$.
The quasi particle energies  $\omega_{i}$'s are given by
 $\omega_1 =\omega +\delta_\xi \quad \omega_2=\omega-\delta_\xi$.
 Here, $\omega=\sqrt{\Delta^2 +\bar\xi^2}$,
$\bar\xi=\bar\epsilon-\bar\nu$ and $\delta _
\xi=(\epsilon_1-\epsilon_2)/2-(\nu_1-\nu_2)/2$. Note that in the
degenerate mass and chemical potential case we shall have the same
quasi particle energies but if the masses and/or the chemical
potentials are different there is possibility of having gapless
modes $\omega_{1} = 0$ ($\omega_{2} = 0$) when
$\omega=-\delta_\xi$ ( $\omega=\delta_\xi$). So, although we shall
have nonzero order parameter $\Delta$, there will be fermionic
zero modes or gapless modes \cite{mishra,igor}. Indeed, we shall
show below that such a situation is possible depending upon the
mismatch in the fermion number densities and the magnitude of
$\Delta$.

 Substituting the solutions for the
ansatz functions in definition for $I_D$, we can have the gap
equation which, however, is divergent. The origin of this
divergence lies in the point-like four fermion interaction which
needs to be regularized. We tackle this  problem by defining the
regularized coupling  as was done in
Refs.\cite{randeria,heiselberg} to access the strong coupling
regime \cite{randeria,bedeque} by subtracting out the zero
temperature and zero density contribution. The regularized gap
equation  is then given by
 \bearr -\frac{1}{g_R} &=& \frac{1}{(2\pi)^3} \int d\zbf k
\left[\frac{1-\theta(-\omega_1)-\theta(-\omega_2)}{2\sqrt{\Delta^2+\bar\xi^2}}
 -\frac{1}{2\bar\epsilon}\right]  \label{gape}
\eearr where $g_R = 2\pi\hbar^2a_{{\rm eff}}/\tilde{m}$. Unless
otherwise stated explicitly, we take  $\hbar=v_F=1$ in our
calculations, where $v_F$ is the Fermi velocity.  Substituting the
expressions for the optimized condensate functions and the
distribution functions, finite part of the thermodynamic potential
Eq. (\ref{omg}), in the zero temperature limit, reduces to \bearr
&\Omega& = \frac{1}{(2\pi)^3}\int d\zbf k\left[\bar\xi-
\sqrt{\Delta^2+\xi^2} \right]  -
\frac{\Delta^2}{g_R}-{g_R}\rho_1\rho_2 \nonumber\\
&+& \frac{1}{(2\pi)^3}\int d\zbf k \left[(\omega+\delta_\xi)
\theta(-\omega_1) +(\omega-\delta_\xi) \theta(-\omega_2)\right] .
\label{thp} \eearr

 \begin{figure}
 \includegraphics[width=3.25in]{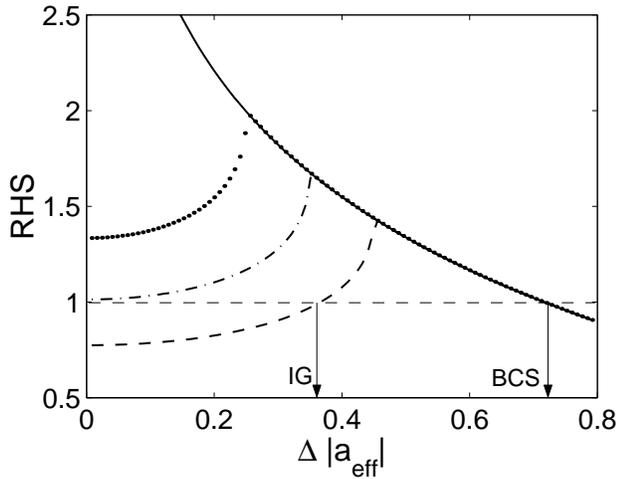}
 \caption{Right hand side (RHS) of gap Eq. (\ref{gape1}) is plotted
 vs $\Delta |a_{{\rm eff}}|$ for different
values of difference in chemical potential $\delta_{\nu}$. The gap
is given by the condition
 RHS=1. The
solid, dotted dashed-dotted and  dashed  lines correspond to
$\delta_\nu  |a_{{\rm eff}}|$=0, 0.25, 0.35 and 0.45,
respectively. For $\delta_\nu  |a_{{\rm eff}}|=0.45$, two
solutions for gap exists. The smaller value corresponds to ``IG
state" and the larger one corresponds to usual
Bardeen-Cooper-Schrieffer state.}
 \label{figdel}
 \end{figure}

In the case of degenerate masses and chemical potentials, the
theta functions will not contribute.  The solution of Eq.
(\ref{gape}) is shown in Fig. \ref{figdel}. In the dilute gas
limit ($k_F |a_{{\rm eff}}| <\!<1$), we have the BCS solution
\cite{heiselberg,randeria} $ \Delta_0 \simeq 8
E_Fe^{-2}\exp[-\pi/(2k_0 |a_{{\rm eff}}|)]$,  where $E_F=
k_0^2/4\tilde m$. This can be seen in Fig.2(a) where $\Delta/E_F$
rises exponentially for small $k_0|a_{eff}|$. At a very high value
of $k_0|a_{{\rm eff}}|$, the gap becomes proportional to the Fermi
energy \cite{heiselberg,randeria}. For the case when $\delta_\xi$
is nonzero, one of the two
 theta functions in Eq. (\ref{gape}) will contribute. To be specific
 let us consider the case when $\delta_\xi$ is negative.
 In that case the contribution from $\theta(-\omega_1)$
will be nonzero in a strip of momenta defined by $k_{{\rm
min}}^2<k^2<k_{{\rm max}}^2$, where $k^2_{{\rm
max,min}}=(m_1\nu_1+m_2\nu_2)\pm
\sqrt{(m_1\nu_1-m_2\nu_2)^2-4\Delta^2m_1m_2}$
\cite{wilczek,bedeque}. The gap equation now becomes
 \bearr &1& =
\frac{|g_R|}{(2\pi)^3}
\int\left[\frac{1}{2\sqrt{\Delta^2+\bar\xi^2}}-\frac{1}{2\bar\epsilon}\right]
d\zbf k -  \frac{g_R}{4\pi^2}\nonumber\\
&\times & \theta\left(\frac{|\delta
k_F^2|}{\sqrt{m_1m_2}}-\Delta\right) \int_{k_{{\rm min}}}^{k_{{\rm
max}}}\frac{k^2}{\sqrt{\Delta^2 +\bar\xi^2}}d k, \label{gape1}
 \eearr
where $\delta k_F^2 = k_{F_1}^2 - k_{F_2}^2$.  For fixed average
chemical potential and for nonzero $\delta_\xi$ there are, in
general, two solutions for the gap  as shown in Fig. 1. Clearly,
when $\delta_\xi=0$, there is only one solution which for the
parameters chosen turns out to be $\Delta=0.71 |a_{{rm
eff}}|^{-1}$. As $\delta_\xi $ is increased, the second integral
in Eq. (\ref{gape1}) starts contributing and the curve can cross
in general the RHS=1 line at two places. The larger $\Delta$ is
the conventional BCS solution. The smaller value of $\Delta$ will
have breached gap character \cite{wil2} and, in this case, the
number densities of the two species will be different. In fact,
the
 number densities of the two species are given as, for $\delta_\xi<0$,
\be \rho_1=\frac{1}{(2\pi)^3}\int\left[\theta(-\omega_1)+
\frac{1}{2}\left(1-\frac{\bar\xi}{\omega}\right)
\left(1-\theta(-\omega_1)\right)\right] d\zbf k \label{rh1}, \ee
 \be
\rho_2=\frac{1}{(2\pi)^3}\int\frac{1}{2}\left(1-\frac{\bar\xi}{\omega}\right
) \left(1-\theta(-\omega_1)\right) d\zbf k. \label{rh2}. \ee

Clearly, without any mismatch in the chemical potential of the two
species, $\omega_1=\omega >0$ and hence the two densities are
equal as the theta functions will not contributes and the
distribution function becomes the usual BCS distribution function.
In presence of mismatch in densities, the $\theta$ function
contributes in the range of momenta between $k_{min}$ and
$k_{max}$. In that region, the particle distribution function
becomes unity for species one and zero for the other species. This
is what is seen in Figs. 2(c) and 2(d). It is easy to see that in
the breached gap phase, the difference in number densities is $
\delta_\rho\equiv\rho_1-\rho_2=(k^3_{{\rm max}}-k^3_{{\rm
min}})/(6\pi^2)$

 \begin{figure}
 \includegraphics[width=3.25in]{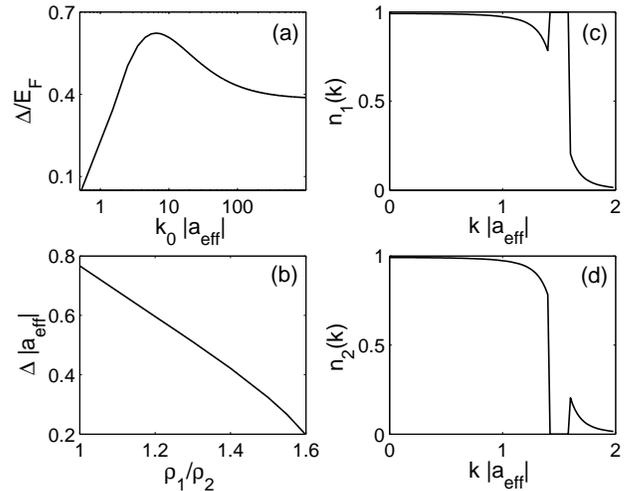}
 \caption{ (a) The gap (in unit of $E_F$) as a function of
 $k_0|a_{{\rm eff}}|$.
 (b) The gap  as a function of
 ratio ($\rho_1/\rho_2$) of the two densities.
  Average  occupation numbers $n_1(k)$ and $n_2(k)$ as functions
  of $k$ are shown in  (c) and (d), respectively, for the
  parameters
$k_0|a_{{\rm eff}}|=1.5$ and $\rho_1/\rho_2=1.4$.}
 \label{figig}
 \end{figure}

For systems where chemical potential is kept fixed, the breached
pair phase will always have higher free energy as compared to the
BCS phase.  However, for systems where the relative densities of
the condensing fermions can be fixed, {\em the breached pairing
state will be the only solution provided the ratio between the
densities is different from unity}. In Fig 2(b), almost linear
behavior of the gap with the density ratio may be noted. Figures
2(c) and (d) show that in a strip of momenta between $k_{min}$ and
$k_{max}$, the momentum distribution resembles that of  normal
fermionic distribution. Below $k_{min}$ and beyond $k_{max}$, the
distribution is of the BCS type. The quasi-particle excitation
energy $\omega_1$ becomes zero at the points $k_{{\rm min}}$ and
$k_{{\rm max}}$ showing gapless behavior. For the parameters
chosen in Figs. 2(c) and (d), $k_{{\rm min}}|a_{{\rm eff}}|=1.42$
and $k_{{\rm max}}|a_{{\rm eff}}|=1.58$, i.e.,  $k_{{\rm
min}}/k_0=0.95$ and $k_{{\rm max}}/k_0=1.05$, $k_0$ being the
average Fermi momentum of the two species.

We thus show that the ansatz state, as defined in Eq. (\ref{bth}),
shows breached pair superfluidity with the excitation energies
becoming zero at  $k_{{\rm min}}$ and $k_{{\rm max}}$ and that
when relative number density is fixed this state is the state
having lower free energy. In this region, only one of the species
is completely occupied and the other is completely empty. Thus, it
is a more general state as compared to Ref.\cite{wilczek} where
the excitation energy was zero at $k_{{\rm min}}$ which was equal
to $p_\Delta$ of Ref. \cite{wilczek}, beyond which one species was
completely occupied and the other was completely empty.

\begin{figure}
 \includegraphics[width=1.5in,height=1.0in]{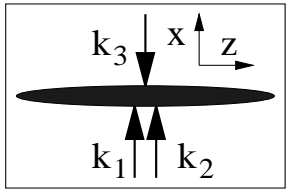}
 \caption{A possible scheme for measuring dynamic structure function
 by SBS.
 The laser photon with momentum
 ${\mathbf k}_1$  is scattered into a
 photon with momentum ${\mathbf k}_3$
 transferring a momentum ${\mathbf q} = {\mathbf k}_1 - {\mathbf
 k}_3$ to the atoms along the x axis, while the laser photon with momentum
 ${\mathbf k}_3$  is scattered into a
 photon with momentum ${\mathbf k}_2$
 transferring a momentum $-{\mathbf q} = {\mathbf k}_3 - {\mathbf
 k}_2$ to the atoms. The frequencies of the three laser beams are
 chosen such that the energy transfer $\delta = (\omega_1 -
 \omega_3) =  (\omega_3 -
 \omega_2) > 0$.  A pair of atoms which may form a Cooper pair having mutually
 opposite momenta are  scattered equally.
 The one-dimensional (along the x direction) momentum and density distribution of the gas can be
 determined
 from the analysis of time of flight images.
 The spectrum (number of scattered atoms versus $\delta$ for different ${\mathbf q}$ values) may reveal the existence
 of the gap and the detailed comparison of the spectra for different relative densities of the two components
 may provide a proof of IG superfluidity.}
 \label{figig}
 \end{figure}

We next discuss the possibility of probing the gap by stimulated
Bragg scattering (SBS) \cite{phonon,bragg} in analogy with Raman
scattering in anisotropic superconductors \cite{klein}. There are
many theoretical proposals \cite{zoller,huletp} for probing atomic
Cooper pairs by resonant \cite{zoller} and off-resonant
\cite{huletp} laser light. It is well established that, in the
case of a Fermi superfluid, there will be no spectrum of density
fluctuation unless energy transferred to the superfluid due to
scattering exceeds $2\Delta$ \cite{klein}. The quantum Monte Carlo
simulation \cite{phand} suggests that, in the large $a_s$ limit,
$\Delta \sim 0.81 (3/5)\epsilon_F \simeq 0.49 \epsilon_F$. In the
recent experiments \cite{sf,salomonnew,molecules} on two-component
$^6$Li atoms in the unitarity-limited regime ($|a_s|k_F
>\!>1$), the typical value
of the Fermi velocity $v_F = \hbar k_F/m \sim$ 15 cm/s. For
two-component $^6$Li atoms, if the Bragg pulses are tuned near an
excited level (say, 2P$_{3/2}$, wavelength $\lambda = 670.776$
nm), the momentum transfer ${\mathbf q} = {\mathbf k}_1 - {\mathbf
k}_3 \simeq 2 k_L \hat{{\mathbf x}}$, where $k_L = 2\pi/\lambda$.
This momentum transfer raises the velocity of the scattered atom
by $2 (\hbar k_L/m) \simeq 20 $ cm/sec which exceeds $v_F$.
Therefore, the scattered atoms can be distinguished in time of
flight images as discussed in the caption of Fig. \ref{figig}.

In conclusion, we have analyzed momentum  distributions and energy
gap for an IG superfluid. Two-component Fermi gases of atoms in
optical traps seem to be promising systems for experimental
exploration for such superfluid state. Fermi atoms in a two-band
optical lattice may also be considered for studying breached pair
superfluidity.

The authors are thankful to  I. Shovkovky for useful
communications and discussions; and  to  M. Randeria for helpful
discussions and bringing to our attention  Ref. \cite{klein}.


\begin{references}

\bibitem{jin} B. DeMacro and D. S. Jin, Science {\bf 285}, 1703
(1999).

\bibitem{huletf} A. G. Truscott, K. E. Strecker, W. I. McAlexander, G.B.
Patridge, and R. G. Hulet, Science {\bf 291}, 2570 (2001).

\bibitem{solomon} F. Schreck {\it et al.}, Phys.Rev.Lett. {\bf 87},
080403 (2001).

\bibitem{thomas} S. R. Granade, M. E. Gehm, K. M. O'Hara, and J. E. Thomas,
Phys.Rev.Lett. {\bf 88}, 120405 (2002).

\bibitem{ketterlef} Z. Hadzibabic {\it et al.},
 Phys.Rev.Lett. {\bf 88}, 160401 (2002).

\bibitem{inguscio} G. Roati, F. Riboli, G. Modungo, and M. Inguscio,
Phys.Rev.Lett. {\bf 89}, 150403 (2002).

\bibitem{regal} C. A. Regal, M. Greiner, and D. S. Jin, Phys. Rev.
Lett. {\bf 92}, 040403 (2004).

\bibitem{coldmol} C. A. Regal {\it et al.}, Nature (London) {\bf 424}, 47
(2003).

\bibitem{salomonnew}  K. E. Strecker {\it et al.}, Phys.
Rev. Lett. {\bf 91}, 080406 (2003); J. Cubizolles {\it et al.},
{\it ibid.} {\bf 91}, 240401 (2003); T. Bourdel {\it et al.}, {\it
ibid.} {\bf 91}, 020402 (2003).

\bibitem{molecules} M. Greiner , C. A. Regal, and D. S. Jin, Nature (London) {\bf 426}, 537
(2003); S. Jochim {\it et al.}, Science {\bf 302}, 2101 (2003); M.
W. Zwierlein {\it et al.},  Phys. Rev. Lett. {\bf 91}, 250401
(2003).

\bibitem{sf} K. M. O'Hara, S. L. Hemmer, M. E. Gehm, S. R. Granade, and
J. E. Thomas, Science {\bf 298}, 2179 (2002).

\bibitem{resonant} M. Holland, S. J. J. M. F. Kokkelmans, M. L.
Chiofalo, and R. Wasler, Phys.Rev.Lett. {\bf 87}, 120406 (2001);
E. Timmermans {\it et al.}, Phys.Lett A {\bf 285}, 228 (2001); Y.
Ohashi and A. Griffin, Phys.Rev.Lett. {\bf 89}, 130402 (2002); W.
Hofstetter {\it et al. }, {\it ibid.} {\bf 89}, 220407 (2002).

\bibitem{wilczek}  W. V. Liu and F. Wilczek
Phys.Rev.Lett. {\bf 90}, 047002 (2003).

\bibitem{sarma} G. Sarma, Phys. Chem. Solid {\bf 24}, 1029 (1963).

\bibitem{wil2} E. Gubankova, W. V. Liu, and F. Wilczek,
Phys. Rev. Lett. {\bf 91}, 032001 (2003).

\bibitem{yip} W. V. Liu and F. Wilczek, e-print cond-matt/0304632;
S. Wu and S. Yip, Phys.Rev. A {\bf 67}, 053603 (2003).

\bibitem{igor} I. Shovkovy and M. Huang, Phys. Lett. B {\bf 564}, 205 (2003).

\bibitem{mishra} A. Mishra and H. Mishra,
Phys Rev D {\bf 69}, 014014 (2004).

\bibitem{rajgopal} M. G. Alford, J. Berges, and K. Rajagopal, Phys. Rev. Lett.
 {\bf 84}, 598 (2000).

\bibitem{rf} S. Gupta {\it et al.}, Science {\bf 300}, 1723
(2003).

\bibitem{zwierlein} M. W. Zwierlein {\it et al.}, Phys. Rev.
Lett. {\bf 91}, 250404 (2003).

\bibitem{stable}  K. M. O'Hara {\it et al.}, Phys.Rev.Lett. {\bf 85}, 2092 (2000).

\bibitem{heiselberg} H. Heiselberg, Phys. Rev. A{\bf 63}, 043606
(2003).

\bibitem{unitary} M. E. Gehm, S. L. Hemmer,  K.M. O'Hara,  and
J. E. Thomas, Phys.Rev.A {\bf 68}, 011603 (2003).

\bibitem{houbiers} M. Houbiers {\it et al.}, Phys. Rev. A {\bf
56}, 4864 (1997); L. Vichi and S. Stringari, Phys. Rev. A {\bf
60}, 4734 (1999); G. M. Bruun, {\it ibid.}  {\bf 66}, 041602
(2002); J.W. Jun, {\it ibid.} {\bf 67}, 043608 (2003).

\bibitem{npa} H. Mishra and J. C. Parikh, Nucl. Phys. A {\bf 679}, 597
(2001).

\bibitem{tfd}
  H.~Umezawa, H.~Matsumoto, and M.~Tachiki {\it Thermofield dynamics
and condensed states} (North-Holland, Amsterdam, 1982); P.
A.~Henning, Phys.~Rep. {\bf 253}, 235 (1995).

\bibitem{amph4} A. Mishra and H. Mishra,
J. Phys. G {\bf 23}, 143 (1997).

\bibitem{randeria} C. A. R. Sa de Melo, M. Randeria, and J.R.
Engelbrecht, Phys. Rev. Lett. {\bf 71}, 3202 (1993).

\bibitem{bedeque} P. F. Bedaque, H. Caldas, and G. Rupak,  Phys. Rev. Lett. {\bf 91}, 247002 (2003).

\bibitem{phonon} D. M. Stamper-Kurn {\it et al.}, Phys. Rev.
Lett. {\bf 83}, 2876 (1999).

\bibitem{bragg} J. Stenger  {\it et al.}, Phys.Rev.Lett. {\bf 82}, 4569
(1999); W. Ketterle and S. Inouye, lanl e-print cond-mat/0101424
(2001).

\bibitem{klein} R. Sooryakumar and M. V. Klein, Phys. Rev. Lett. {\bf
45}, 660 (1980); M. V. Klein and S. B. Dierker, Phys. Rev. B {\bf
29}, 4976 (1984).

\bibitem{zoller} P. T\"{o}rm\"{a} and P. Zoller, Phys. Rev. Lett. {\bf 85}, 487 (2000);
G. M. Bruun {\it et al.}, Phys. Rev. A {\bf 64}, 033609 (2001).

\bibitem{huletp} W. Zhang, C. A. Sackett, and R. G. Hulet, Phys. Rev. A {\bf 60}, 504 (1999);
J. Ruostekoski, {\it ibid.} {\bf 60}, 1775 (1999); M. Rodriguez
and P. T\"{o}rm\"{a}, {\it ibid.} {\bf 66}, 033601 (2002);

\bibitem{phand} J. Carlson {\it et al.}, Phys. Rev. Lett. {\bf
91}, 050401 (2003).

\end{references}
\end{document}